\documentclass[usenatbib]{mnras}
\usepackage{aas_macros}
\usepackage{amsmath}
\usepackage{amssymb}
\usepackage{color}
\usepackage{epsfig}
\usepackage{float}
\usepackage{graphicx}
\usepackage{latexsym}
\usepackage{morefloats}
\usepackage{natbib}
\usepackage{subfigure}
\usepackage{times}
\usepackage[normalem]{ulem}
\usepackage{hyperref}

\newcommand{\plotone}[1]{\resizebox{0.95\hsize}{!}{\includegraphics{#1}}}

\newcommand{\Msun}{{~ M_\odot}}

\newcommand{\mpch}{~h^{-1}\rm Mpc}

\def\gsim { \lower .75ex \hbox{$\sim$} \llap{\raise .27ex \hbox{$>$}}}
\def\lsim { \lower .75ex \hbox{$\sim$} \llap{\raise .27ex \hbox{$<$}}}

\newcommand{\elucid}{\textsc{elucid}}

\newcommand{\reffig}[1]{Fig. \ref{#1}}

\definecolor{blue-violet}{rgb}{0.54, 0.17, 0.89}
\definecolor{purple}{rgb}{0.5, 0., 0.5}

\title[LOCUSTS]
{Screening maps of the local Universe I -- Methodology}
\author[Shao et al.]
{\parbox{\textwidth}{Shi Shao$^{1}$\thanks{E-mail: shi.shao@durham.ac.uk}, Baojiu Li$^1$, Marius Cautun$^{1,2}$, Huiyuan Wang$^3$ and Jie Wang$^{4,5}$ \vspace{.20cm}}\\
$^1$Institute for Computational Cosmology, Department of Physics, Durham University, South Road Durham DH1 3LE, UK \\
$^2$ Leiden Observatory, Leiden University, PO Box 9513, NL-2300 RA Leiden, the Netherlands \\
$^3$Key Laboratory for Research in Galaxies \& Cosmology, Department of Astronomy, University of Science \& Technology of China, Hefei, Anhui
230026, China\\
$^4$Key Laboratory for Computational Astrophysics, National Astronomical Observatories, Chinese Academy of Sciences, Beijing, 100012, China\\
$^5$University of Chinese Academy of Sciences, 19 A Yuquan Rd, Shijingshan District, Beijing, 100049, China
}

\pubyear{2018}

\begin{document}
\label{firstpage}
\pagerange{\pageref{firstpage}--\pageref{lastpage}}
\maketitle

\begin{abstract}
We introduce the {\sc loc}al {\sc u}niverse {\sc s}creening {\sc t}est {\sc s}uite ({\sc locusts}) project, an effort to create `screening maps' in the nearby Universe to identify regions in our neighbourhood which are screened, i.e., regions where deviations from General Relativity (GR) are suppressed, in various modified gravity (MG) models. In these models, deviations from the GR force law are often stronger for smaller astrophysical objects, making them ideal test beds of gravity in the local Universe. However, the actual behaviour of the modified gravity force also depends on the environment of the objects, and to make accurate predictions one has to take the latter into account. This can be done approximately using luminous objects in the local Universe as tracers of the underlying dark matter field. Here, we propose a new approach that takes advantage of state-of-the-art Bayesian reconstruction of the mass distribution in the Universe, which allows us to solve the modified gravity equations and predict the screening effect more accurately. This is the first of a series of works, in which we present our methodology and some qualitative results of screening for a specific MG model, $f(R)$ gravity. Applications to test models using observations and extensions to other classes of models will be studied in future works. The screening maps of this work can be found at this link\footnotemark.
\end{abstract}

\begin{keywords}
cosmology: theory -- gravitation -- methods: numerical
\end{keywords}

\footnotetext{\url{http://icc.dur.ac.uk/~sshao/locusts/}}
\section{Introduction}
\label{sect:intro}

In recent years, modified gravity (MG) theories \citep{Clifton:2011jh,Joyce:2014kja,Koyama:2015vza,Koyama:2018} have been an active field of research in theoretical, observational and computational cosmology. One of the primary motivations for studying such models is to find alternative models to explain the accelerated cosmic expansion \citep{Riess:1998cb,Perlmutter:1998np}, that avoid the theoretical difficulties in the standard $\Lambda$-cold-dark-matter ($\Lambda$CDM) paradigm. {Other motivations for MG theories include} attempts to find a more complete theory of gravity than General Relativity (GR) and to develop new ways to test the accuracy of GR; the latter is of particular interest since cosmological observations have entered the precision era, and started to allow accurate tests of gravity on length and energy scales vastly different from where GR has been conventionally validated \citep[e.g.,][]{Will:2014kxa}.

Being a long-range force, gravity acts on all length scales from sub-atomic to cosmological. Therefore, a deviation from GR's prescription can in principle be measured on all these scales. Hence, although many of the MG models are originally proposed to tackle a cosmological problem, they can be tested in a huge array of environments or regimes, from laboratory experiments \citep[see][for a recent review]{Brax:2018}, to Solar system and astrophysical objects \citep[see][for a recent review]{Sakstein:2018}, and to observations at cosmological distances \citep[see][for some recent reviews]{Koyama:2015vza,Heymans:2018,Cataneo:2018,Cai:2018}.

The requirement that any new theory of gravity must preserve the success of GR on small length scales has important implications on both theories and observations. Theoretically, one is confined to {\it viable} MG models, i.e., those that behave sufficiently closely to GR in environments such as the Solar System. One way to achieve this is through a screening mechanism \citep[e.g.,][]{Khoury:2010xi}, by which modifications to the GR force law are suppressed in places of deep {gravitational} potential or {in regions characterised by} large gradients and/or {by large} Laplacians of the potential (like in the Solar system). Observationally, this implies that viable MG models must pass local tests of gravity by design, and thus we may need to turn to astrophysical and cosmological probes for complementary and potentially more stringent tests. The latter has been possible because cosmology concerns {typically} environments with shallow gravitational potentials {or small values of its derivatives}, where order unity deviations from GR {can} occur. MG theories {are characterised by a variety of} screening mechanisms, {which means} that a given probe could have very different constraining power for different models. Therefore, it is sensible to explore a wide range of potential cosmological and astrophysical probes. For example, for the popular $f(R)$ gravity model, in which the deviation from GR is controlled by a model parameter $f_{R0}$ (see more details below), the strongest constraints on $f_{R0}$ from cosmology {suggest} $|f_{R0}|\lesssim10^{-6}$ \citep[e.g.,][]{He:2018oai,Leo:2019}, while the astrophysical constraints are claimed to be stronger \citep[e.g.,][]{Jain:2012tn,Sakstein:2014nfa}.

Even if one is interested in astrophysical constraints, it is often not sufficient to focus only on individual astrophysical objects. This is because, as we have mentioned above, the deviation from GR in many MG models is dependent on not just the astrophysical objects themselves but also the properties of their environments. A dwarf galaxy, for example, can be unscreened (i.e., it experiences a modified gravitational force) if placed in a low-density environment for a specific $f(R)$ model, but the same galaxy may well be screened (i.e., the deviation from GR is efficiently suppressed) if moved to dense environments such as close to a large galaxy cluster. In other words, screening is a nonlinear phenomenon, and the behaviour of (modified) gravity on small scales can not be cleanly disentangled from its behaviour on much larger scales. As a result, the precise knowledge of the total matter distribution in a large region (the environment) is necessary to accurately predict how a modified gravity model would affect the observational properties of an astrophysical object. Not knowing the former could introduce a uncontrolled systematical uncertainty to astrophysical tests of gravity.

Fortunately, observations of the local Universe have now become good enough for us to `reconstruct' the relevant environmental properties needed to understand the screening. The first attempt of making use of such vital information was by \cite{Cabre:2012}, who estimated {at the position of each observed galaxy} the Newtonian potential, $\Phi_{\rm env}$, -- which determines the screening efficiency for $f(R)$ gravity -- from all other neighbouring galaxies:
\begin{equation}
    \Phi_{\rm env} = \sum\frac{GM_i}{r_i}.
\end{equation}

A similar but more sophisticated approach was taken by \cite{Desmond:2017ctk}, who considered also $\nabla\Phi_{\rm env}$ and $\nabla^2\Phi_{\rm env}$, which are quantities controlling the efficiency of other screening mechanisms than the chameleon mechanism exploited by $f(R)$ gravity. More effort was also devoted to obtaining the underlying mass distribution. In \cite{Cabre:2012} only the galaxies detected by the Sloan Digital Sky Survey ({\sc sdss}) were utilised to reconstruct $\Phi_{\rm env}$, while \cite{Desmond:2017ctk} also included the contributions from (i) invisible dark matter haloes -- haloes which do not host a galaxy -- by using a simulation calibration, and (ii) the underlying total matter field (not necessarily in resolved haloes) at $z=0$, as obtained by a Bayesian density reconstruction technique \citep{Lavaux:2015tsa}. The results of the works are 3D maps of the local Universe, which contain values of $\Phi_{\rm env}$, $\nabla\Phi_{\rm env}$ and $\nabla^2\Phi_{\rm env}$: these are called screening maps as these quantities determine the screening properties of the leading MG models as a function of location.

In this paper we introduce a new approach to obtain screening maps. Our approach also makes use of the reconstructed total matter field from the observed galaxy catalogues in the local Universe. However, instead of using this density field to calculate quantities such as $\Phi$ and $\nabla\Phi$, we directly use that to solve for the dynamical fields which are responsible for the modification of gravity (and for screening). The main motivation is that, while the above quantities qualitatively determine the efficiency of screening, the quantitative calculation is much more involving: as an example, in Vainshtein-type models it is not $\nabla^2\Phi$, but $\nabla^2\phi$ and $\nabla^i\nabla^j\phi\nabla_i\nabla_j\phi$, where $\phi$ is a scalar field propagating the modified gravity force, that determines the screening, and this is further complicated by the complex cosmic web. This approach, dubbed {\sc loc}al {\sc u}niverse {\sc s}creening {\sc Test} {\sc S}uite, or {\sc locusts}, solves $\phi$ using the reconstructed density field by employing routines of the MG numerical simulation code {\sc ecosmog} \citep{Li:2011vk,Li:2013nua,Li:2013tda}. This therefore requires the MG model to be clearly specified, and the study will be on a model-by-model basis. On the other hand, because there is only one observed local Universe, the underlying model of gravity -- whichever it is -- must reproduce the observationally-inferred matter density field. In particular, simulations of different gravity models should produce this same matter density field at $z\sim0$, perhaps starting from different initial conditions. As a result, we only need to run one single $\Lambda$CDM simulation and output the matter field at various snapshots, and then the modified gravity routine in {\sc ecosmog} can be used to calculate the screening properties of the model in these snapshots. This is much faster than full MG cosmological simulations, so that we can easily repeat the calculation for hundreds or even thousands (therefore the name {\sc locusts}) of MG models that densely sample the model and parameter space. Another possibility enabled by this approach is the study of the time evolution of the screening map, which can be obtained by running the MG solver in {\sc ecosmog} on several close output snapshots and then doing a finite difference. 

In this paper we describe the methodology of the {\sc locusts} simulations, and show the screening maps and some other physical quantities to demonstrate how it works. We do these using a specific MG model -- chameleon $f(R)$ gravity -- as an example, leaving the application of the method in astrophysical tests and extensions of it to include larger coverage of the local Universe and of more MG models into future works. 

The layout of this paper is as follows. Section~\ref{sect:simul} briefly reviews the chameleon $f(R)$ gravity theory and the simulations used in this work. Section~\ref{sect:results} presents our results, including visualisations of the simulated haloes and scalar field compared with the observed distributions of galaxies and galaxy groups, some simple statistics of the behaviour of the fifth force, and detailed properties of the COMA cluster. Finally, we conclude with a short summary and discussion in Section~\ref{sect:conclusions}.

\section{Methods and simulations}
\label{sect:simul}

\subsection{Constrained simulations of the local Universe}
\label{subsect:constrained_sims}

We make use of a constrained-realisation {\it N}-body simulation (labelled as CS) performed as part of the \elucid{} project \citep{Wang2014, Wang2016}. The goal of the project is to reproduce the evolution history of our Local Universe by using the reconstructed initial density field from the observed galaxy catalogue. Here, we briefly summarise the reconstruction method as follows. First, a halo-based group catalogue is constructed from the SDSS DR7 galaxy catalogue with their positions and velocities having been corrected to ``real space". Then, a present-day density field is built according to the obtained halo catalogue. Finally, using the Hamiltonian Markov Chain Monte Carlos (HMC) algorithm with particle mesh dynamics, the initial condition is reconstructed from the present-day density field. For a more detailed description, we refer the reader to \citet{Wang2016}. The method can effectively trace the $z=0$ massive haloes ($\gsim 10^{13.5} \Msun$) back to their initial condition, such that the reconstructed initial condition of our Local Universe can be used to study the evolution history of individual galaxy clusters and other cosmic web environments (see \reffig{fig:map}).

The initial condition reconstructed above, which is used in this work, features a periodic cubic box with a side length $500h^{-1}$Mpc and $1024^3$ dark matter particles. The mass of each simulation particle is {$8.3 \times 10^9h^{-1}\Msun$. The cosmological parameters are adopted from the best-fit WMAP5 cosmology \citep{Dunkley2009}: $\Omega_{\rm m}=0.258, \Omega_\Lambda=0.742, h=0.72,\sigma_8=0.8$ and $n_{\rm s}=0.96$.

\subsection{The theoretical model}
\label{subsect:model}

In this work we focus on a particular class of modified gravity models, $f(R)$ gravity, which is an extension to standard GR by replacing the Ricci scalar $R$ in the Einstein-Hilbert action of gravity with an algebraic function of $R$:
\begin{equation}\label{eq:S-f(R)}
\mathcal{S}_{\rm EH} = \int \mathrm{d}^4x\sqrt{-g} \frac{1}{16\pi G} \left[R + f(R)\right],
\end{equation}
where $G$ is Newton's constant and $g$ is the determinant of the metric $g_{\mu\nu}$, with $\mu,\nu=0,1,2,3$. 

The modified Einstein equation can be obtained by varying the action, Eq.~\eqref{eq:S-f(R)}, with respect to the metric $g_{\mu\nu}$, to obtain
\begin{equation}\label{eq:M-EQ}
G_{\mu\nu} + f_R R_{\mu\nu} - g_{\mu\nu} \left[\frac{1}{2}f(R) - \Box f_R \right] - \nabla_\mu \nabla_\nu f_R = 8\pi G T^{\rm m}_{\mu\nu}\,,
\end{equation}
in which $G_{\mu\nu} \equiv R_{\mu\nu} - \frac{1}{2}g_{\mu\nu} R$ {denotes the usual} Einstein tensor, $\nabla_\mu$ is the covariant
derivative compatible with $g_{\mu\nu}$, $\Box \equiv \nabla^\mu \nabla_\mu$ is the d'Alambertian, and $T^{\rm m}_{\mu \nu}$ is the energy-momentum tensor for matter. The quantity $f_R$ in this equation is an extra degree of freedom (a scalar field) of this model, defined by
\begin{equation}\label{eq:sc}
f_R \equiv \frac{\mathrm{d}f(R)}{\mathrm{d}R},
\end{equation}
whose equation of motion can be obtained by taking the trace of Eqn.~\eqref{eq:M-EQ}:
\begin{equation}\label{eq:f_R}
\Box f_R = \frac{1}{3}\left[R - f_R R + 2f(R) + 8\pi G\rho_{\rm m}\right],
\end{equation}
where $\rho_{\rm m}$ is the density of non-relativistic matter. Therefore, the scalar field $f_R$ satisfies a second-order field equation of motion; this means that the modified Einstein equation, \eqref{eq:M-EQ}, which contains fourth-order derivatives of $g_{\mu\nu}$, can be rewritten as a standard second-order Einstein equation with a scalar field.

To investigate the evolution of cosmic structures in the Newtonian regime, we derive the perturbation equations in the Newtonian gauge on a flat Friedmann-Robertson-Walker (FRW) background:
\begin{equation}\label{eq:FRW}
\mathrm{d}s^2 = (1+2\Psi) \mathrm{d}t^2 - a^2(t)(1-2\Phi)\delta_{ij}\mathrm{d}x^i \mathrm{d}x^j,
\end{equation}
in which $\Phi=\Phi({\bf x},t)$ and $\Psi=\Phi({\bf x},t)$ are the gravitational potentials, which are functions of the physical time $t$ and the comoving coordinates ${\bf }=\{x^i\}$; $\delta_{ij}$ is the 3D spatial metric, and $a(t)$ is the scale factor, which is normalised to $a(t_0) = a_0 = 1$ at the present day (a subscript $_0$ denotes the current value of a quantity throughout this paper, unless otherwise stated). In the quasi-static and weak field limits, the system of equations, \eqref{eq:M-EQ} and \eqref{eq:f_R}, can be simplified respectively to:
\begin{eqnarray}\label{eq:Phi}
\nabla^2 \Phi &=& \frac{16}{3}\pi Ga^2\delta\rho_{\rm m} + \frac{1}{6} a^2\delta R,\\
\label{eq:fR}
\nabla^2 f_R &=& -\frac{1}{3}a^2[\delta R + 8\pi G\delta \rho_{\rm m}],
\end{eqnarray}
in which $\nabla^2$ denotes the 3D Laplacian operator, and the density and curvature perturbations are defined respectively as
$\delta \rho_{\rm m} \equiv \rho_{\rm m} - \bar{\rho}_{\rm m}$ and $\delta R \equiv R(f_R) - \bar{R}$; an overbar is used to denote the background value of a quantity. Eq.~(\ref{eq:Phi}) can be recast in a new form:
\begin{equation}\label{eq:Phi2}
\nabla^2\Phi = 4\pi Ga^2\delta\rho_{\rm m} - \frac{1}{2}\nabla^2f_R.
\end{equation}
It can be seen clearly that the second term of the right-hand side of Eq.~\eqref{eq:Phi2} represents a modification to the standard Poisson equation, and we can define $\Phi\equiv\Phi_{\rm GR}-\frac{1}{2}f_R$, where $\Phi_{\rm GR}$ is the Newtonian potential in GR, and $-\frac{1}{2}f_R$ can be identified as the potential of an additional force -- the so-called fifth force, which is propagated by the scalar field $f_R$ -- between matter particles. The fifth force is not detected in solar system or laboratory tests of gravity \citep[][]{Will:2014kxa}, and these experimental tests place strong constraints on models like this. 

To close Eqs.~(\ref{eq:Phi},\ref{eq:fR}), one needs the relationship between $f_R$ and $R$ such that $\delta R$ can be expressed as a function of the scalar field as $f_R$: $\delta R\left(f_R\right)$. This can be done by specifying the functional form of $f(R)$, which satisfies the requirement that the resulting $f_R$ is a monotonic function of $R$. If $f(R)$ is a slowly-varying function of $R$, i.e., $|f_R|\ll1$, the model has two desirable features:
\begin{itemize}
\item the terms involving $f_R$ in Eq.~\eqref{eq:M-EQ} can be neglected to a good approximation, reducing the Einstein equation to
\begin{equation}
G_{\mu\nu} - \frac{1}{2}g_{\mu\nu}f(R) \approx 8\pi GT_{\mu\nu}.
\end{equation}
If one further approximates $f(R)\approx-2\Lambda$ (recall that $f(R)$ is taken to be nearly constant), with $\Lambda$ being the cosmological constant, then the background expansion history of this model can be made close to that of $\Lambda$CDM. {In fact,} with suitable choices of $f(R)$ the background expansion histories in the two models can be made exactly identical \citep[][]{He:2013}.
\item if $|f_R|\ll1$, one can have $\nabla^2f_R\sim0$ and consequently from Eq.~\eqref{eq:Phi2} we can see that the standard Poisson equation in GR is recovered. If this happens at least in high-density regions, it implies that the fifth force is suppressed in such regions, which can make the model compatible with current local tests of GR.
\end{itemize}

The suppression of the fifth force in the limit $|f_R|\ll1$ is the result of a suitable choice of $f(R)$; it is a dynamical effect called the screening mechanism. $f(R)$ gravity is a representative example of a wider class of models, called the chameleon model \citep{Khoury:2003rn}, in which the suppression (or screening) of the fifth force works as following: the scalar field $f_R$, which propagates the fifth force between matter particles, satisfies Eq.~\eqref{eq:fR}, which can be rewritten as
\begin{equation}
\nabla^2f_R+\frac{\partial V_{\rm eff}\left(f_R\right)}{\partial f_R} = 0,
\end{equation}
where $V_{\rm eff}\left(f_R\right)$ is an effective potential of the scalar field, given by
\begin{equation}\label{eq:fR2}
\frac{\partial V_{\rm eff}\left(f_R\right)}{\partial f_R} =\frac{1}{3}a^2[\delta R\left(f_R\right) + 8\pi G\delta \rho_{\rm m}].
\end{equation}
The potential $V_{\rm eff}$ characterises the interactions of the scalar field with itself (the first term on the right side of Eq.~\eqref{eq:fR2}) and matter (the second term). For a choice of $f(R)$ such that $V_{\rm eff}\left(f_R\right)$ has a global minimum at $f_R=f_{R,\rm min}$ and $f_{R,{\rm min}}\rightarrow0$ as $\delta\rho_{\rm m}\rightarrow\infty$, the fifth force can be suppressed in high-density regions as desired, therefore evading the stringent local constraints on it. Because the behaviour of the fifth force is dependent on the environmental density, the screening mechanism is called the chameleon mechanism. In regions where $|\delta\rho_{\rm m}|\ll1$, on the other hand, the curvature perturbation $|\delta R|\ll1$ and so from Eqs.~(\ref{eq:Phi},\ref{eq:fR}) one can derive that
\begin{equation}
\nabla^2\Phi \approx \frac{16}{3}\pi G\delta\rho_{\rm m} = \frac{4}{3}\nabla^2\Phi_{\rm GR},
\end{equation}
which means that gravity is enhanced by a factor of $1/3$ -- an effect that is potentially testable using cosmological observations.

The actual behaviour of the fifth force in $f(R)$ gravity is more complicated that the above intuitive picture, and an accurate solution has to be made by numerically solving Eq.~\eqref{eq:fR} given a matter configuration. In this context, to solve for $f_R$ at a given position we need its solution in the neighbourhood as the boundary condition -- in other words, to know for certain whether a given cosmological object, such as a star or galaxy, is screened, we need to solve Eq.~(\ref{eq:fR}) in a large region encompassing this object, and the solution in that region in turn depends on further nearby regions, and so on. In this picture, screening of the fifth force for an object can be achieved in two ways:
\begin{itemize}
\item self screening: if the objective is massive enough, it alone can make $|f_R|$ small inside and/or nearby, therefore screening the fifth force it feels;
\item environmental screening: if the object is not massive enough to self screen, but lives near some much larger objects, then $|f_R|\ll1$ can still be satisfied inside and/or near it, causing a suppression of the fifth force it experiences.
\end{itemize}
To use astrophysical objects in the local Universe to test the fifth force, then, we cannot reliably treat those objects as isolated bodies living on the cosmological background, but have to take into account their larger-scale environments. For this reason a constrained realisation simulation as described in Section \ref{subsect:constrained_sims}, where the matter distribution mimics that in the real observed local Universe, is ideal as it offers a way to more realistically model the effect of environments in the chameleon screening. 

In this work we shall set up the general strategy to carry out constrained realisation simulations in modified gravity models, and present some first results to show how it works. We leave detailed analyses of these simulations that lead to constraints on model parameters to future works. For concreteness, we use the $f(R)$ model proposed by \citet[HS;][]{Hu:2007nk} as example. This model is given by specifying
\begin{equation}\label{eq:f(R)}
f(R) = -m^2 \frac{c_1}{c_2} \frac{(-R/m^2)^n}{(-R/m^2)^n + 1}\,,
\end{equation}
where $m^2 \equiv 8\pi G \bar{\rho}_{{\rm m}0}/3 = H^2_0 \Omega_{\rm m}$ is a parameter of mass dimension 2, $\Omega_{\rm m}$ the density parameter for non-relativistic matter, $H_0$ the present-day value of the Hubble expansion rate, and $n$, $c_1$ and $c_2$ are dimensionless model parameters. The scalar field, Eq.~\eqref{eq:sc}, takes the following form:
\begin{equation}\label{eq:sc1}
f_R = -\frac{c_1}{c^2_2} \frac{n(-R/m^2)^{n-1}}{[(-R/m^2)^n + 1]^2}.
\end{equation}

To see whether this model can have a background expansion history close to that of standard $\Lambda$CDM, let us consider a $\Lambda$CDM model with $\Omega_{\rm m}\approx0.3$ and $\Omega_\Lambda=1-\Omega_{\rm m}\approx0.7$, for which we find $|\bar{R}| \approx 40 m^2 \gg m^2$, and therefore
\begin{equation}\label{eq:sc2}
f_R \approx -n\frac{c_1}{c^2_2} \left(\frac{m^2}{-R} \right)^{n+1}.
\end{equation}
For $n\sim1$ and $c_1/c_2^2\lesssim1$, we then have $|f_R\left(\bar{R}\right)|\ll1$, which is the condition by which the background expansion history is close to $\Lambda$CDM, and
\begin{equation}\label{eq:c1c2}
f(R) \approx -m^2\frac{c_1}{c_2} \approx -2\Lambda \Rightarrow \frac{c_1}{c_2} = 6\frac{\Omega_\Lambda}{\Omega_{\rm m}}.
\end{equation} 


Therefore, once we have specified an (approximate) $\Lambda$CDM background history (by which $c_1/c_2$ is fixed), the HS $f(R)$ model then has two free parameters, $n$ and $c_1/c^2_2$. The latter is related to the present-day value of the background scalaron, $f_{R0}$,
\begin{equation}\label{eq:c12}
\frac{c_1}{c^2_2} = -\frac{1}{n} \left[ 3\left(1 + 4\frac{\Omega_\Lambda}{\Omega_{\rm m}} \right) \right]^{n+1} f_{R0}\,.
\end{equation}
The choice of $f_{R0}$ and $n$ fully determines the model.

\subsection{The {\sc locusts} simulations}

In this subsection we introduce the {\sc locusts} simulation suite and briefly describe the simulation technique used.

The {\sc locusts} simulations are a suite of simulations of various modified gravity models, all starting from an identical initial condition, which itself is obtained as described in Section \ref{subsect:constrained_sims}. Therefore, they are the first attempt to realistically simulate our local Universe in the context of modified gravity. In particular, one of the primary objectives of {\sc locusts} is to obtain screening maps, namely a map to show the screening properties at different spatial locations in the local Universe. As stated in the introduction, such screening maps can provide vital information for both cosmological and astrophysical tests of gravity. 

While the basic idea is general, in this work we focus on the chameleon $f(R)$ gravity model described in Section \ref{subsect:model} as explicit example. In particular, we shall specialise to the case of $n=1$, and run simulations for 20 different values of $|f_{R0}|$, ranging from $10^{-7}$ to $10^{-6}$. This parameter range is still compatible with the currently most stringent constraints on $f_{R0}$ from cosmological observations \citep[see, e.g.,][]{Cataneo:2014kaa,Liu:2016xes,Peirone:2016wca}. 

The chameleon $f(R)$ simulations used in this work have been done using the {\sc ecosmog} \citep{Li:2011vk} code, which is a modified version of the publicly available $N$-body and hydrodynamical simulation code {\sc ramses} \citep{Teyssier:2001cp}. This is a particle-mesh code employing the adaptive-mesh refinement technique to achieve high force resolution in dense regions, and parallelised using message passing interface. {\sc ecosmog} extends {\sc ramses} by solving the nonlinear field equations which arise from various modified gravity models numerically by the multigrid relaxation method. For details about the implementation in different classes of models, see \citet{Li:2011vk,Li:2013nua,Li:2013tda} and references therein. We use an optimised version of {\sc ecosmog} for the Hu-Sawicki $f(R)$ model, as described in \cite{Bose:2016wms}, which is based on a more efficient algorithm to solve the $f(R)$ field equation.

Even with the algorithm optimisation from \cite{Bose:2016wms}, running a suite of $>\mathcal{O}(20)$ simulations with different $f(R)$ gravity parameters is still computationally expensive for the resolution and particle number used in {\sc locusts}. Fortunately, as explained in the introduction, the idea behind {\sc locusts} does not require us to run full simulations of modified gravity, but only needs one simulation to $z=0$ which provides a mock universe with a underlying matter density field. This underlying density field must be as close to the observationally-inferred density field in the local Universe as possible, and any gravity model should reproduce this same underlying density field. This might be achieved by tuning the initial conditions of the simulations in different models, but the details are not our concern here. Apparently, the simplest way to achieve this is to only run the full simulation (from $z_{\rm ini}=80$ to $z=0$) in the $\Lambda$CDM model, while for the $f(R)$ models we simply run the {\sc ecosmog} for a few steps -- respectively on the particle snapshots of the $\Lambda$CDM simulation at various redshifts -- to calculate the behaviour of the scalar field and the fifth force at the redshifts of interest to us. Put in other words, the particle evolution in the {\sc locusts} simulations is done using Newtonian gravity, while the evaluation of the screening map is done using the complete modified gravity solver.

As mentioned above, the evolution of particle positions and the calculation of the scalar field and screening properties are both performed using {\sc ecosmog}, which is based on the {\sc ramses} code. As a rough estimate of the level to which we can trust the simulation density field (i.e., the typical difference between different simulation codes at our resolution), in Fig.~\ref{fig:pk_check} we have compared the matter power spectra at $z=0$ predicted by {\sc ecosmog} and the {\sc gadget}-2 code \citep{Springel:2005mi}. We can see there is good agreement -- within $1\%$ for $k<3~h$Mpc$^{-1}$ and $4\%$ for $k<6~h$Mpc$^{-1}$. The difference at small scales ($\lesssim\mathcal{O}(1)h$Mpc$^{-1}$) is expected to be much smaller than the typical uncertainty in the density reconstruction.

\begin{figure}
	\plotone{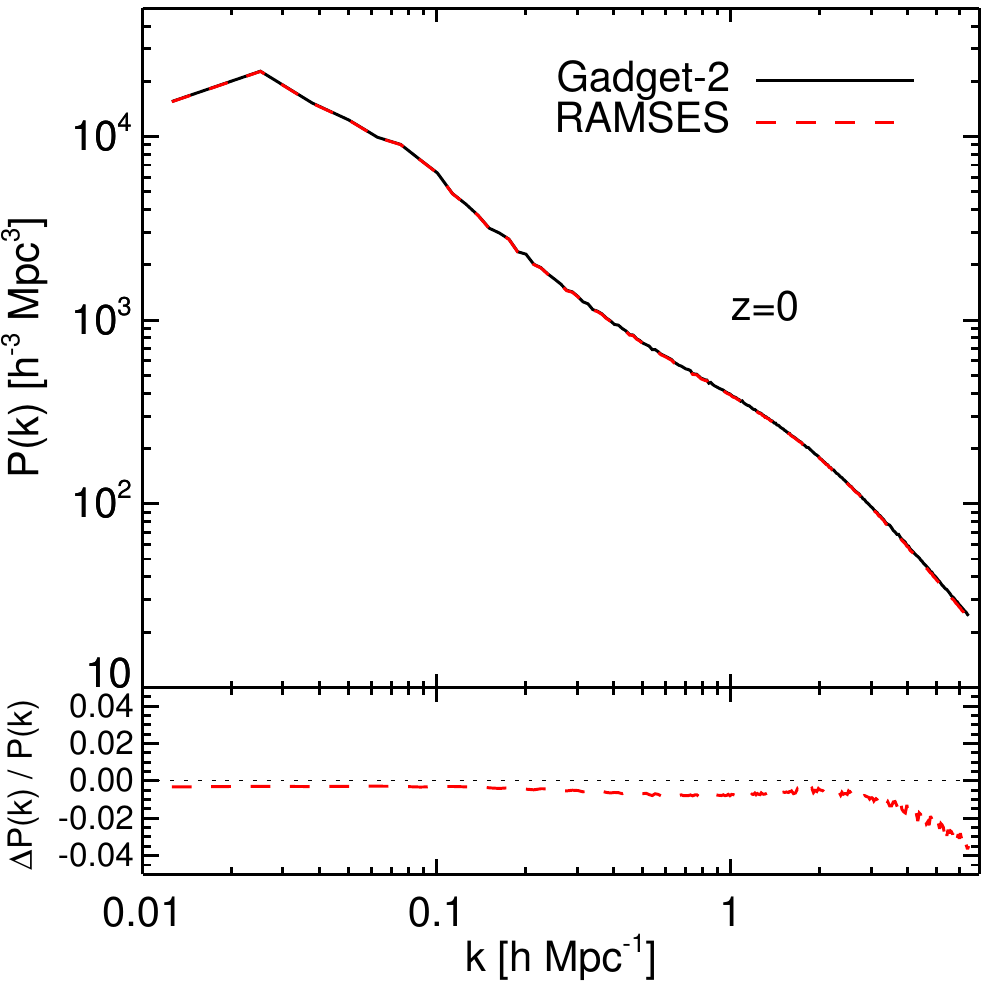}
	\caption{Comparison of the $z=0$ power spectrum between the two constrained simulations (labelled as Gadget-2 and RAMSES), which were run using the {\sc gadget}-2 (solid) and the {\sc ramses} (dashed) codes, respectively. Both simulations have the same initial condition. The bottom panel shows the residual difference between the two simulations.}
    \label{fig:pk_check}
\end{figure}

\begin{figure}
	\plotone{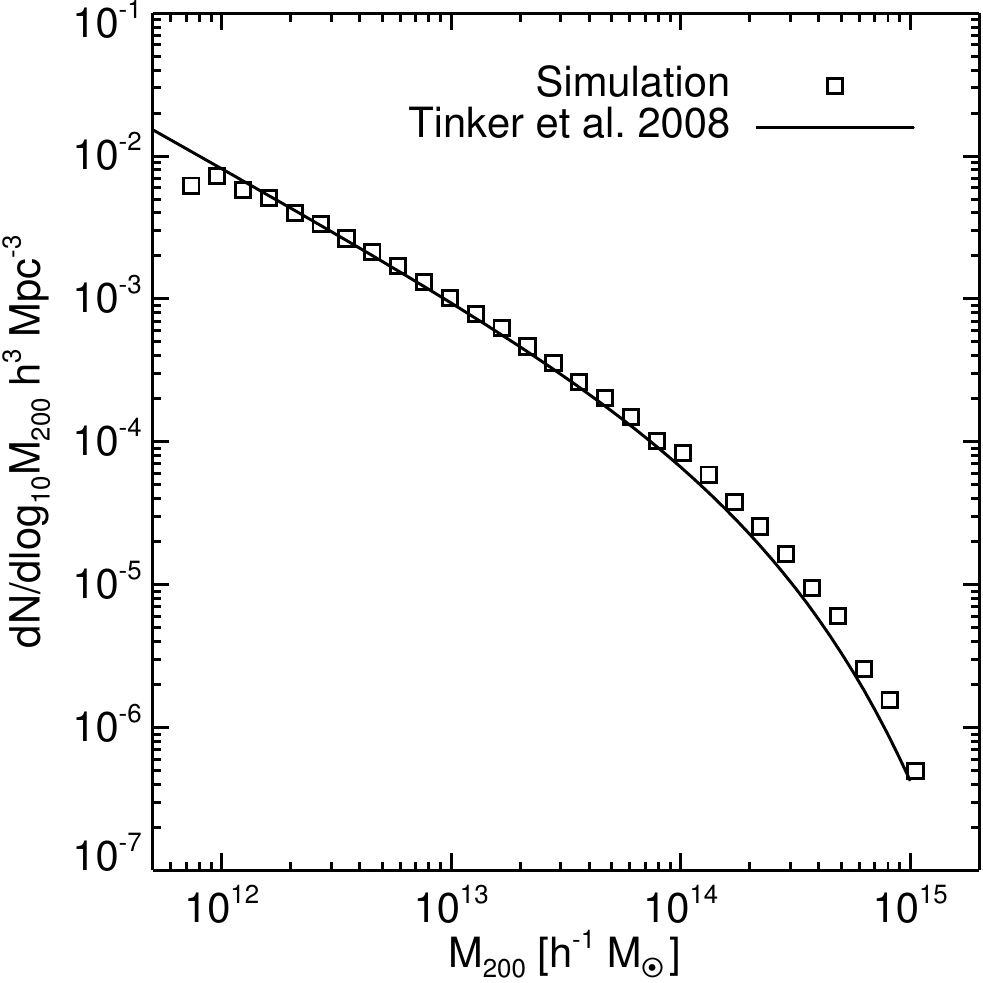}
	\caption{The halo mass function of the $z=0$ GR simulations. The solid line shows the \citet{Tinker_2008} mass function.} 
    \label{fig:mf_halo}
\end{figure}

As another sanity check, in Fig.~\ref{fig:mf_halo} we plot the halo mass function from the $\Lambda$CDM simulation at $z=0$ (squares) compared with the \citet{Tinker_2008} fitting formula (solid line). The halo catalogues in this and other figures of this paper are identified using the phase-space friends-of-friends halo finder {\sc rockstar} \citep{rockstar}, and the halo mass $M_{200}$ denotes the mass within $R_{200}$, the radius within which the average density is 200 times that of the critical density of the Universe at the halo redshift, $\rho_{\rm crit}(z)$. The simulation output agrees well with \citet{Tinker_2008} apart from the high-mass end, and at $M_{200}\lesssim10^{12}h^{-1}\Msun$ (which correspond to haloes with $\lesssim100$ particles, for which the mass function becomes incomplete due to the low resolution).

\section{Results}
\label{sect:results}

This section contains the main results of this work. We start with some visualisation and general properties of the fifth force throughout the simulation box, then move on to study statistical properties of the screening maps and the screening around prominent structures in the local Universe, such as the Coma cluster and the SDSS Great Wall.

\subsection{Visualisation}

\begin{figure*}
	\plotone{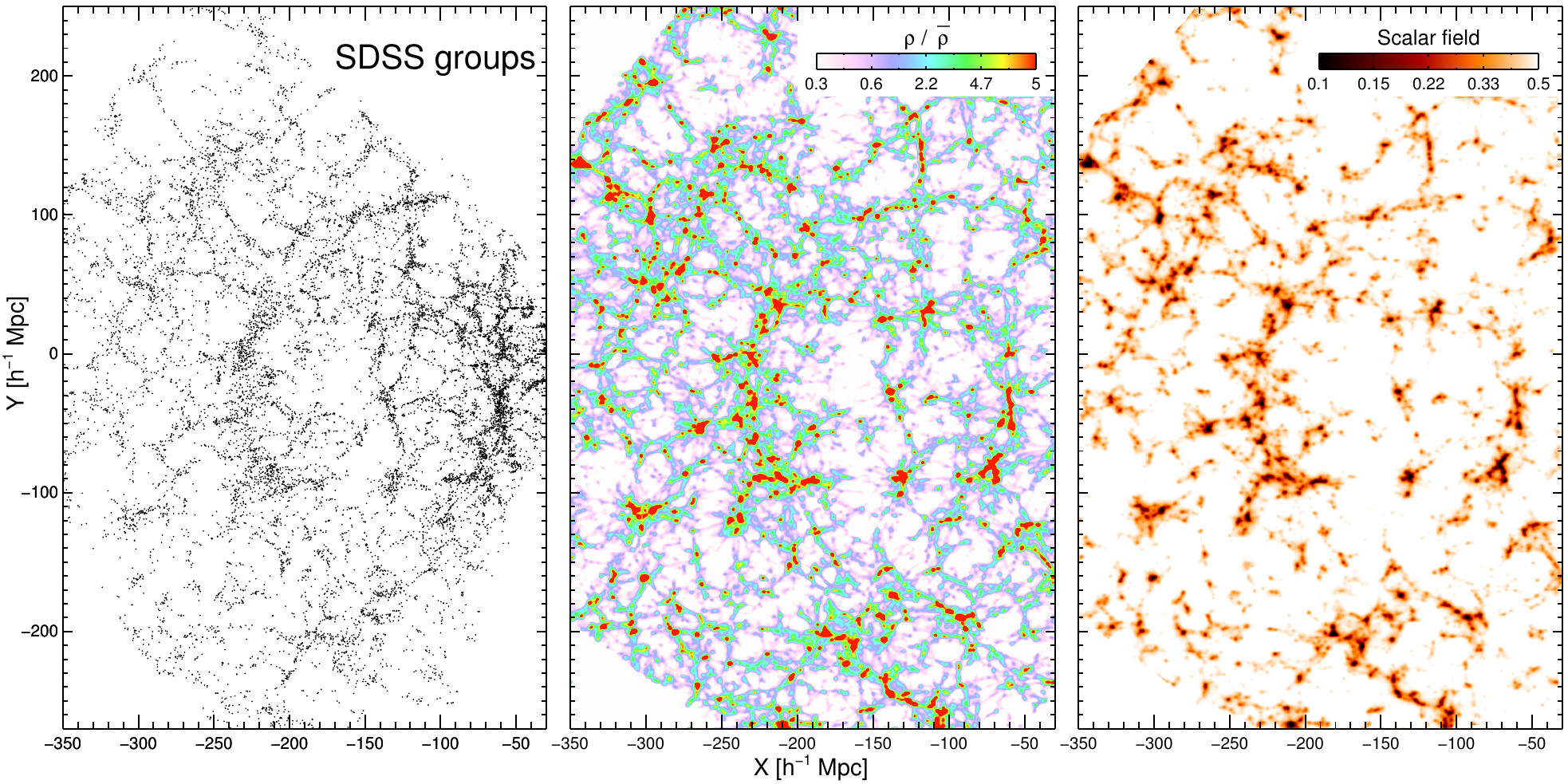}
    \caption{A visualisation comparison of the observed local Universe and the one reproduced in our constrained simulation. {\it Left Panel}: galaxy group distribution as observed by SDSS in a slice with thickness of $10\mpch{}$. {\it Middle Panel}: the dark matter distribution as predicted by our cosmological simulation constrained to reproduce the local Universe. {\it Right Panel}: The scalar field in the same region as the middle panel, for the model with $|f_{R0}|=10^{-6}$.}
    \label{fig:map}
\end{figure*}

Fig.~\ref{fig:map} is the visual comparison of a slice taken from the SDSS group catalogue (left panel, in which groups are shown as black dots) with an extraction of the simulation box that is supposed to represent the same region (middle and right panes); the middle panel shows the dark matter density field in the region, while the right panel shows the corresponding scalar field configuration for $f_R$. Both simulation results are at $z=0$, and for the right panel a particular $f(R)$ model with $|f_{R0}|=10^{-6}$ is shown for illustration purpose.

We see from Fig \ref{fig:map} that the constrained simulation has successfully reproduced the large-scale structures observed from the SDSS catalogue, noticeably the filamentary patterns on scales of tens of Megaparsecs and above. In particular, from the dark matter distribution we can see clearly the SDSS Great Wall found at $X=-230\mpch{}$ and extending in the vertical coordinate from $-100$ to $50\mpch$ \citep{Goot_05}. The scalar field $f_R$, as shown in the right panel, behaves as expected from the chameleon screening mechanism: its value is closer to $0$ near clusters and filaments, while approaching the background value $f_{R0}$ further away from these structures. In particular, we note that deep inside void regions the scalar field is nearly uniform, suggesting that the fifth force, which is the gradient of the scalar field, is weak there\footnote{However, the fifth force can still be significant near (often small) matter clumps inside these voids, and we shall return to this point later below.}.

To better compare our constrained simulation with the observational data, we zoom-in on a small region centred on the Great Wall. The results are shown in Fig.~\ref{fig:galaxy}, where the upper left panel is an enlarged view of the matter density field from the central panel of Fig.~\ref{fig:map} using the same colour bar.

Dark matter haloes identified in the constrained simulation are shown as black open circles in the lower left panel of Fig.~\ref{fig:galaxy}, where the radius of each circle is proportional to the mass of the halo it represents. 
Overplotted on top are the SDSS galaxy groups which are shown as red filled circles. We find a very good agreement between the positions of simulated haloes and those of SDSS groups, although some outliers do exist. This comparison represents a beautiful illustration of how well the constrained simulation reproduces the large-scale distribution of galaxies.

In the two panels on the right-hand side of Fig.~\ref{fig:galaxy} we show the SDSS groups and galaxies overplotted on screening maps for the same zoomed-in region. The coloured map in the upper right panel shows again the scalar field, where we can see more clearly that the scalar field closely traces matter distribution and is nearly homogeneous in low-density regions. The blue and red dots represent SDSS blue and red galaxies respectively in this panel, and the latter also trace well the simulation matter distribution. This suggests that we can use the simulated screening map to predict the scalar field value and fifth force ratio at the positions of the observed objects. In the lower-right panel we show this for groups (filled circles whose sizes indicate the masses of the groups they represent) and red galaxies (dots) -- here the colour is used to illustrate the fifth force ratio at the positions of the groups and galaxies, and we can see that the objects are more screened in dense regions than in underdense regions.

\begin{figure*}
	\plotone{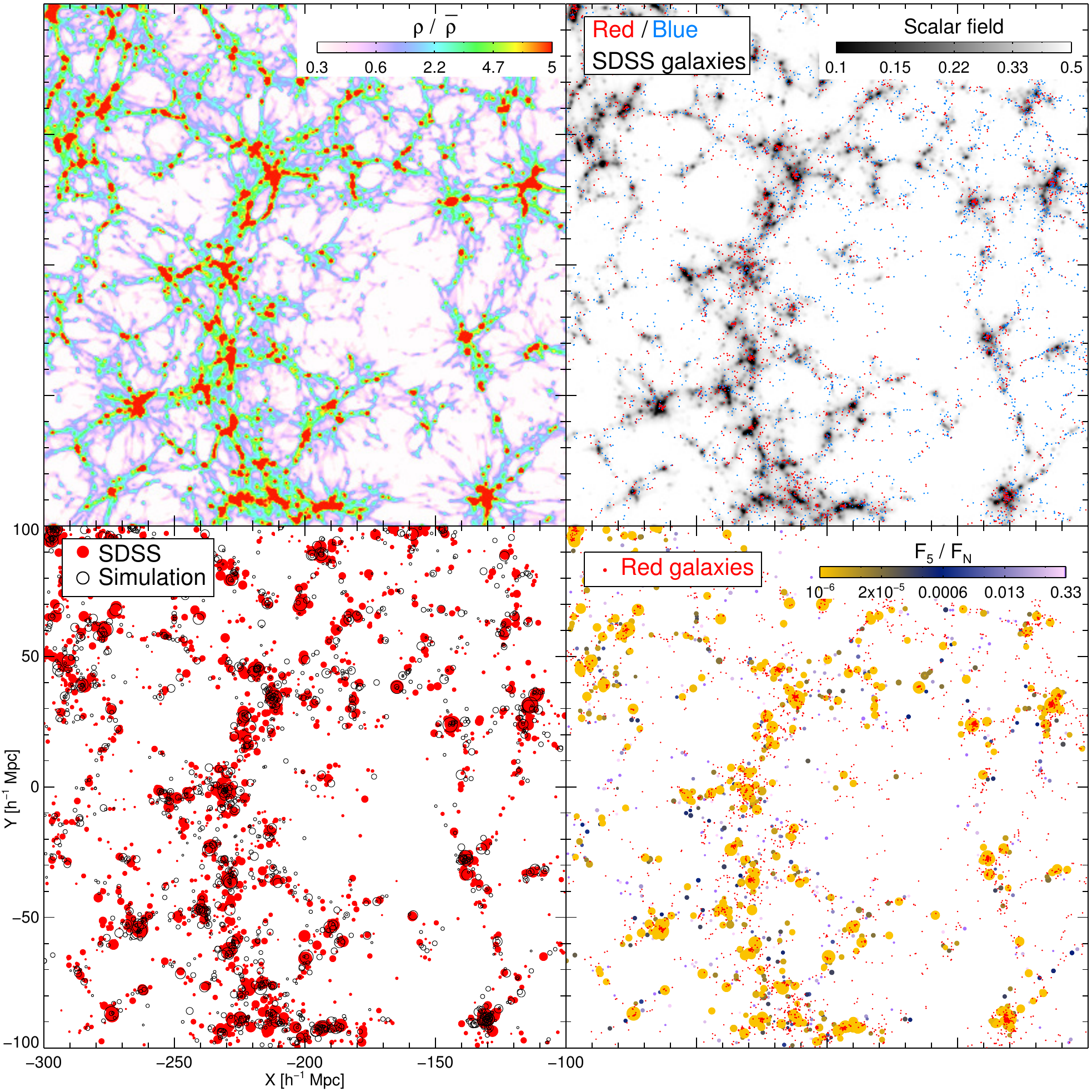}
	\caption{{\it Top-left}: the dark matter density field in a selected region near the SDSS Great Wall, where red colours show high-density regions and white colours show low-density regions (see colour bar). {\it Top-right}: the scalar field in the same selected region, where dark grey and white show the screened (small field) and unscreened (large field) regimes. SDSS red and blue galaxies are overplotted as the red and blue points, respectively. {\it Bottom-left}: SDSS galaxy groups (red filled circles) and dark matter haloes from our simulation (black empty circles) in the same selected region, with the sizes of the circles representing the mass of the groups or haloes. {\it Bottom-right:} SDSS groups (filled circles with varying sizes as the bottom left panel) and red galaxies (dots) in the same selected region, with the colour showing the ratio between the magnitudes of the fifth and Newtonian gravity forces at the positions of the object, as predicted by our simulation. Colour bars used in each panel are shown individually.
	}
    \label{fig:galaxy}
\end{figure*}

\subsection{Generic behaviours of the fifth force}

\begin{figure*}
	\plotone{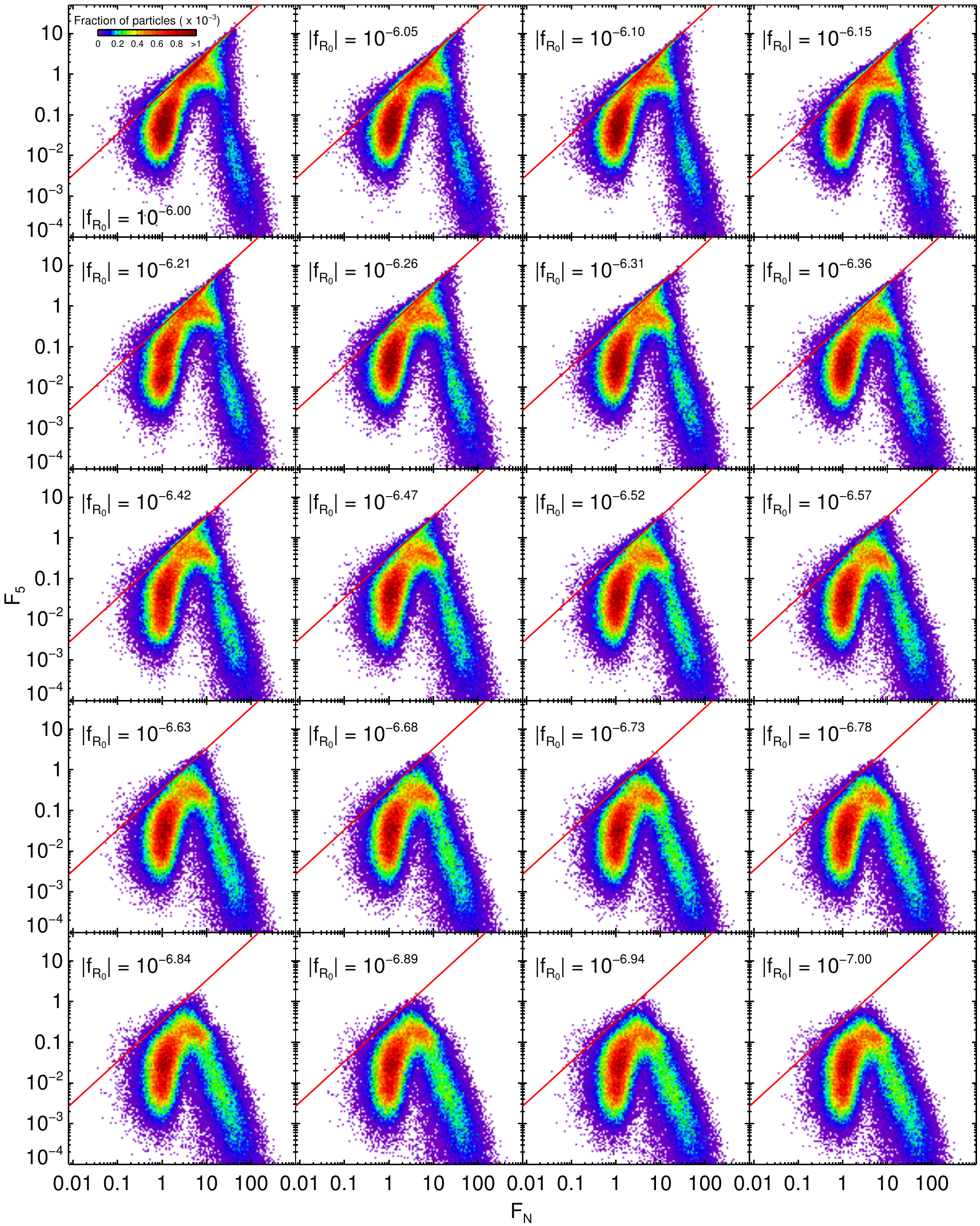}
	\caption{Fifth force against standard gravity force for a subset of particles randomly selected	from the z=0 simulation box. The colour indicates the number of particles, with red (purple) showing that many (few) particles have the given force values. The panels correspond to different MG models where $|{f_R}_0|$ varies from $10^{-6}$ (top-left panel) to $10^{-7}$ (bottom-right panel). The red line shows the analytical prediction in which the fifth force is $1/3$ times the magnitude of standard Newtonian gravity. The force values are in code unit, $L_{\rm box}H_0^2$, where $L_{\rm box}=500\mpch$ is the side length of the simulation box and $H_0=72 \rm km/s/Mpc$ is the present-day hubble constant. All results are from the $z=0$ snapshot.
	}
	\label{fig:force_ratio_z0}
\end{figure*}

\begin{figure*}
	\plotone{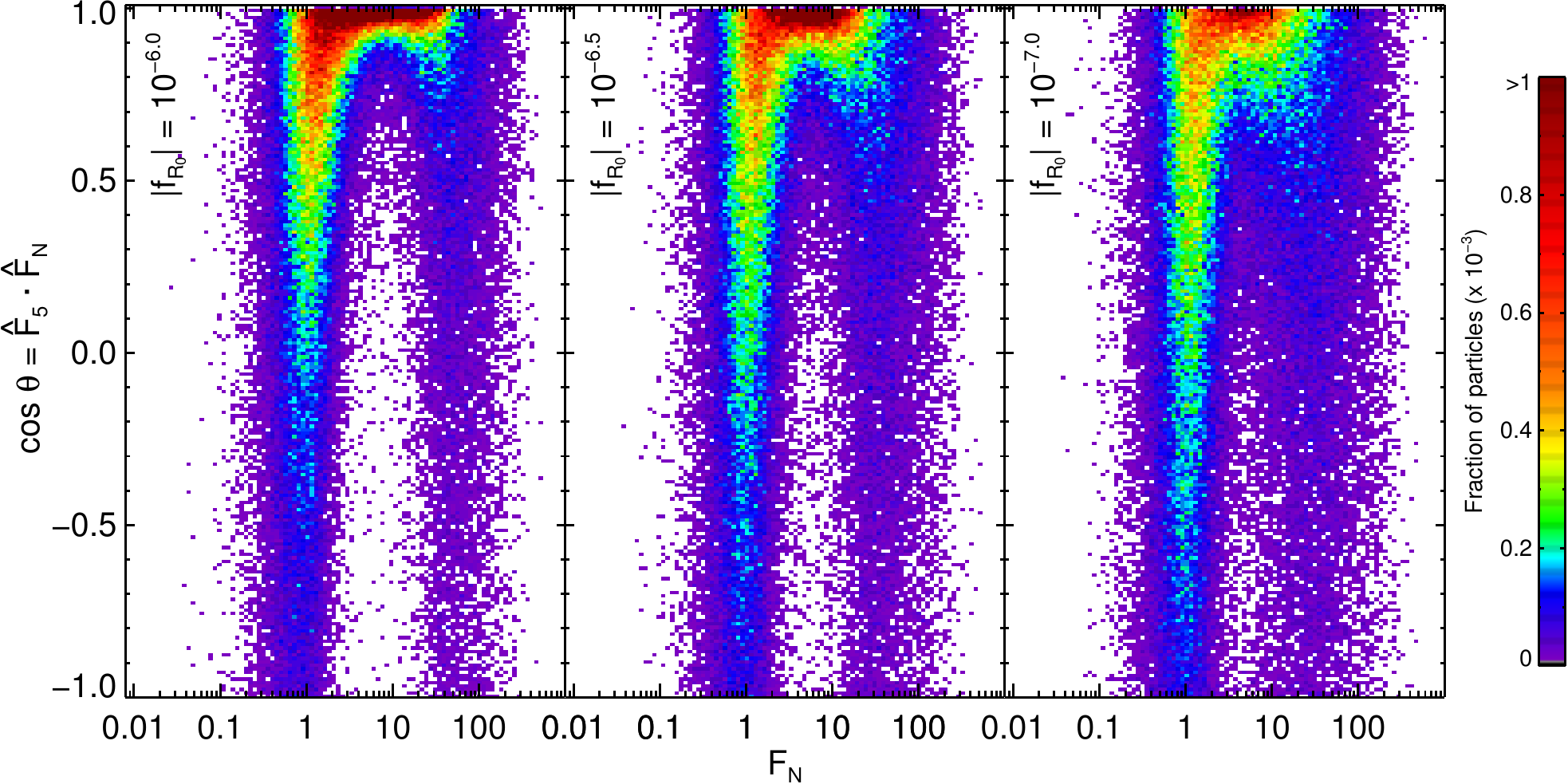}
	\caption{Similar to \reffig{fig:force_ratio_z0} but with the vertical axis showing $\cos\theta$, where $\theta$ is the alignment angle, between the fifth and standard gravity forces, $\cos\theta\equiv{\bf F}_5\cdot{\bf F}_{\rm N}/(|F_5|\cdot|F_{\rm N}|)$. From left to right, we only show $|{f_R}_0|=10^{-6}, 10^{-6.5}$, and $10^{-7}$ respectively.}
    \label{fig:force_angle_z0}
\end{figure*}

Before quantifying the fifth force effects, let us present some results of the general behaviour of the fifth-force-to-standard-gravity ratio across the whole simulation volume. In Fig.~\ref{fig:force_ratio_z0}, we have shown this force ratio at the positions of $10^5$ particles randomly selected from the simulation box, where each dot represents the measured value at a simulation particle. The different panels are for different $|f_{R0}|$ values, starting from the least screened case with $10^{-6}$ at the upper left and ending at the most strongly screened case with $10^{-7}$ at the lower right. The colour indicates the frequency that particles appear with given standard gravity (horizontal axis) and fifth force (vertical axis) values. Comparing amongst the different panels and comparing simulation results with analytical linear perturbation prediction (red solid line), we observe the following features:
\begin{itemize}
    \item In high-density regions, where the magnitude of the standard gravity force is large, the fifth force is generally strongly screened, and the points are well below the red line, which represents the case that the fifth force has $1/3$ of the strength of standard gravity.
    \item In the regime of intermediate magnitudes of standard gravity, representative for smaller haloes and filaments, the fifth force ratio agrees with linear theory prediction well for the weakly screened models. However, as $|f_{R0}|$ decreases, stronger screening shows up even in this regime; for example, in the last row, we can see clearly that the red dots are well below the red solid line.
    \item In the regime of weak standard gravity, i.e, the left end of each panel, which is representative of void regions, the fifth force ratio falls below the red solid line again. This is because the fifth force, unlike standard Newtonian gravity, is a short-ranged force that decays exponentially beyond the Compton wavelength of the scalar field. This implies that the standard gravity exerted by particles outside the void regions can reach the inner part of these voids, while the fifth force cannot, leading to a suppressed force ratio between the latter and the former. This can also be understood through the observation that in void regions, e.g., Fig.~\ref{fig:galaxy}, the scalar field is nearly homogeneous and so the fifth force becomes weak. Note that the deviation from linear (anaytic) prediction of the fifth-force-to-standard-gravity ratio does not happen in Vainshtein screening models, e.g., Fig.~3 of \cite{Falck:2014jwa} since there the fifth force is long range.
\end{itemize}

Figure \ref{fig:force_angle_z0} is similar to Figure \ref{fig:force_ratio_z0}, but instead of the force ratio, it shows the cosine of the angle $\theta$ between the fifth and standard gravity forces. If linear theory works perfectly, the fifth force should be $1/3$ of the strength of standard gravity and the directions of the two forces would be the same. While this is the case for most particles from the intermediate gravity regime (the red and orange regions), we can see that for the strong and weak gravity regimes this is not true. The reason is the same as for the behaviour of the force ratio shown in Fig.~\ref{fig:force_ratio_z0}, namely in regions of deep Newtonian potential (and therefore strong standard gravity) the fifth force is suppressed by the chameleon screening mechanism, while inside voids the fifth force from matter in surrounding regions suffers from the Yukawa exponential decay -- because of that, and given the irregular matter distribution, the fifth force and Newtonian gravity on a particle in a weak-gravity region can receive contributions from different neighbouring particles as vector additions, and so they do not necessarily have the same direction; this is particular true because the fifth force has much smaller amplitude (note that even in the weak-gravity regions there are still particles, so the fifth force is not exactly zero).

\begin{figure*}
	\plotone{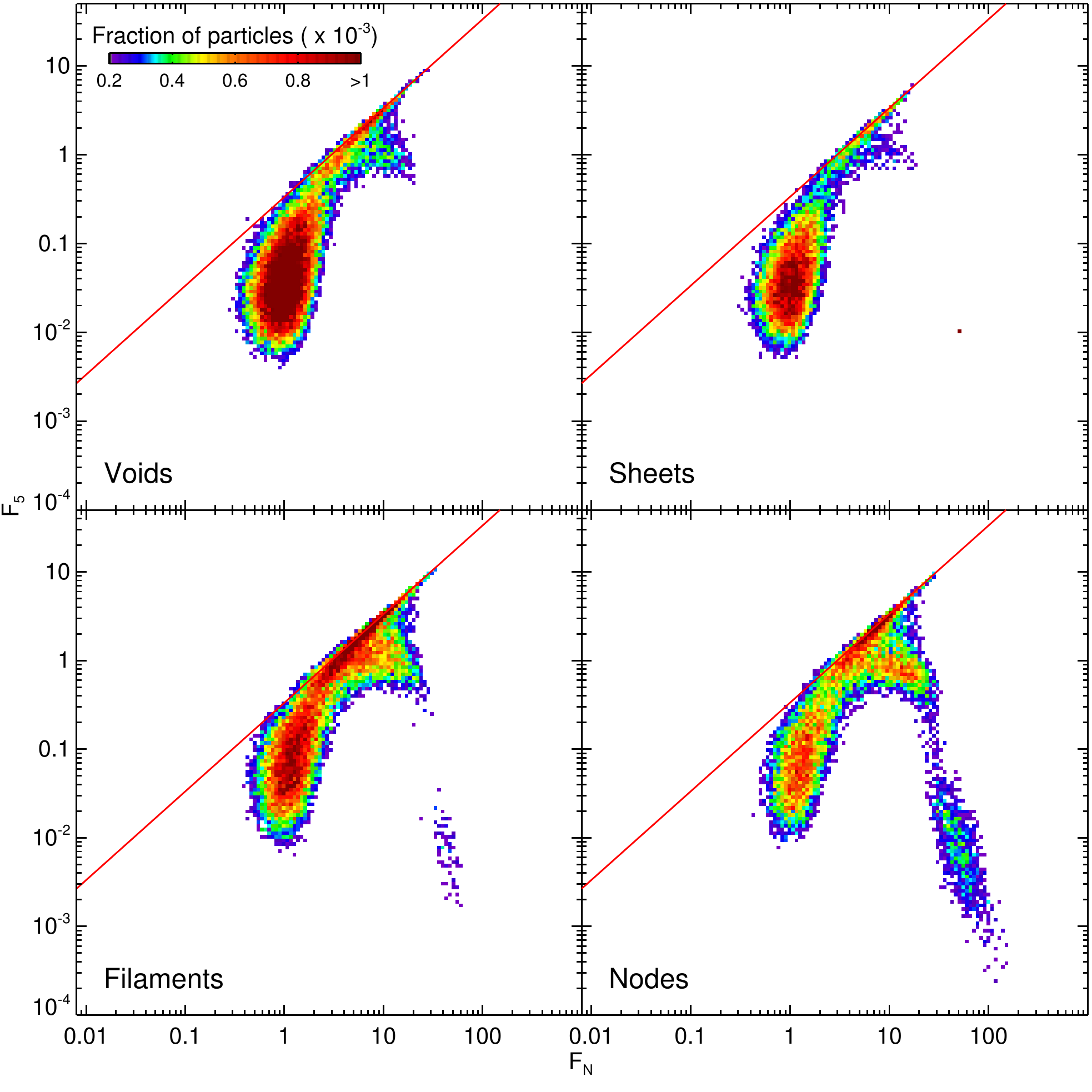}
	\caption{ The dependence of the fifth force against standard gravity for DM particles split according to their cosmic web environment. We show results for void (top-left), sheet (top-right), filament (bottom-left) and node (bottom-right) environments identified using the \textsc{nexus+} method.
	}
	\label{fig:morphology}
\end{figure*}

Finally, we are interested to check how the force ratio depends on large-scale environment. For this we used the \textsc{nexus+} method \citep{Cautun2013} to identify the various cosmic web environments: nodes, filaments, sheets and voids. The nodes correspond to the densest regions, filaments to 1D linear structures, sheets to 2D wall-like planar densities and voids to underdense regions. These morphological environments have been found by first using the Delaunay Tessellation Field Estimator \citep{Schaap2000,Cautun2011} to calculate the density field on a regular grid with a $1\mpch$ grid spacing. Then, \textsc{nexus+} calculates the eigenvalues, $\lambda_i$ with $\lambda_1\leq\lambda_2\leq\lambda_3$, of the Hessian matrix of the smoothed density field, which are used to classify the web environments. The exact procedure is based on some rather complex functions of the Hessian eigenvalues, however the result can be qualitatively understood as: nodes correspond to regions with $\lambda_1\approx\lambda_2\approx\lambda_3<0$, filaments to regions with $\lambda_1\approx\lambda_2<0$ and $\lambda_2\ll\lambda_3$, sheets to $\lambda_1<0$ and $\lambda_1\ll\lambda_2$, and voids to everything else. For a detailed comparison of the \textsc{nexus+} technique to other web finders, please see \citet{Libeskind2018}.

The resulting cosmic web is dominated in terms of volume by voids, which occupy ${\sim}80\%$ of the volume but contain only ${\sim}15\%$ of the total mass budget. In terms of mass, the filaments are the most important environment, containing over half of the mass budget but filling only $6\%$ of the cosmic volume \citep{Cautun2014}. Most of the massive haloes, with $M_{200}\gtrsim5\times10^{13}h^{-1}\Msun$, are found in nodes, while filaments contain the majority of lower mass haloes with mass $M_{200}\gtrsim10^{11}h^{-1}\Msun$ \citep{GaneshaiahVeena2018}. In contrast, sheets and especially voids correspond to below average densities and are mostly devoid of haloes with masses above $10^{12}h^{-1}\Msun$. This means that the majority of bright galaxies, that is with stellar masses above $10^9h^{-1}\Msun$, are found in either the filaments or nodes of the cosmic web \citep{GaneshaiahVeena2019}.

Figure \ref{fig:morphology} shows the same fifth-force-to-standard-gravity ratio as in Figure~\ref{fig:force_ratio_z0}, but for particles found in voids (upper left), sheets (upper right), filaments (lower left) and nodes (lower right). To increase the clarity of the plots, we have only shown the results for $|f_{R0}|=10^{-6}$ and neglected pixels which represent particles that are smaller than $0.2$ thousandth of the total particle number. The overall behaviour is similar to what Figure \ref{fig:force_ratio_z0} shows, but there is also a clear distinction between the various web environments. For example, the long drop-off tail with small force ratio but strong standard gravity forces seen in Figure \ref{fig:force_ratio_z0} is mainly due to particles from nodes (high-density environments), while the drop-off from the analytical line at weak standard gravity forces is dominated by low-density environments such as voids and sheets, as explained above.

\subsection{The Coma Cluster}

The constrained initial condition used in our simulations has a limited volume, with objects such as the Local Group and Virgo Cluster not included. Therefore, here we select the object corresponding to the Coma cluster in our simulation volume, to illustrate the behaviour of the modified gravity force in massive objects.

Coma is a cluster at a distance of about $100$ Mpc from us, with over 1000 member galaxies and a total mass of $\sim10^{15}M_\odot$. The dark matter halo we identify from our simulation as the counterpart of Coma\footnote{In what follows we shall refer to both the real Coma cluster and the counterpart dark matter halo from our simulation as `Coma'; the context should make clear what we mean.} is found to have a mass of $M_{\rm 200}=7.7\times10^{14}h^{-1}\Msun$ and halo radius $R_{\rm 200}=1.5h^{-1}$Mpc. As a first visual inspection, in Figure~\ref{fig:coma_density} we show the projected density in a $40\times40$ $\left(h^{-1}{\rm Mpc}\right)^2$ field of view centred around the Coma halo, with a projection depth of $5h^{-1}$Mpc. On top of this, the observed Coma member galaxy groups are also shown as black open circles. We can see that the galaxy groups broadly follow the same clustering pattern of high-density regions in the projected map.

\begin{figure*}
	\plotone{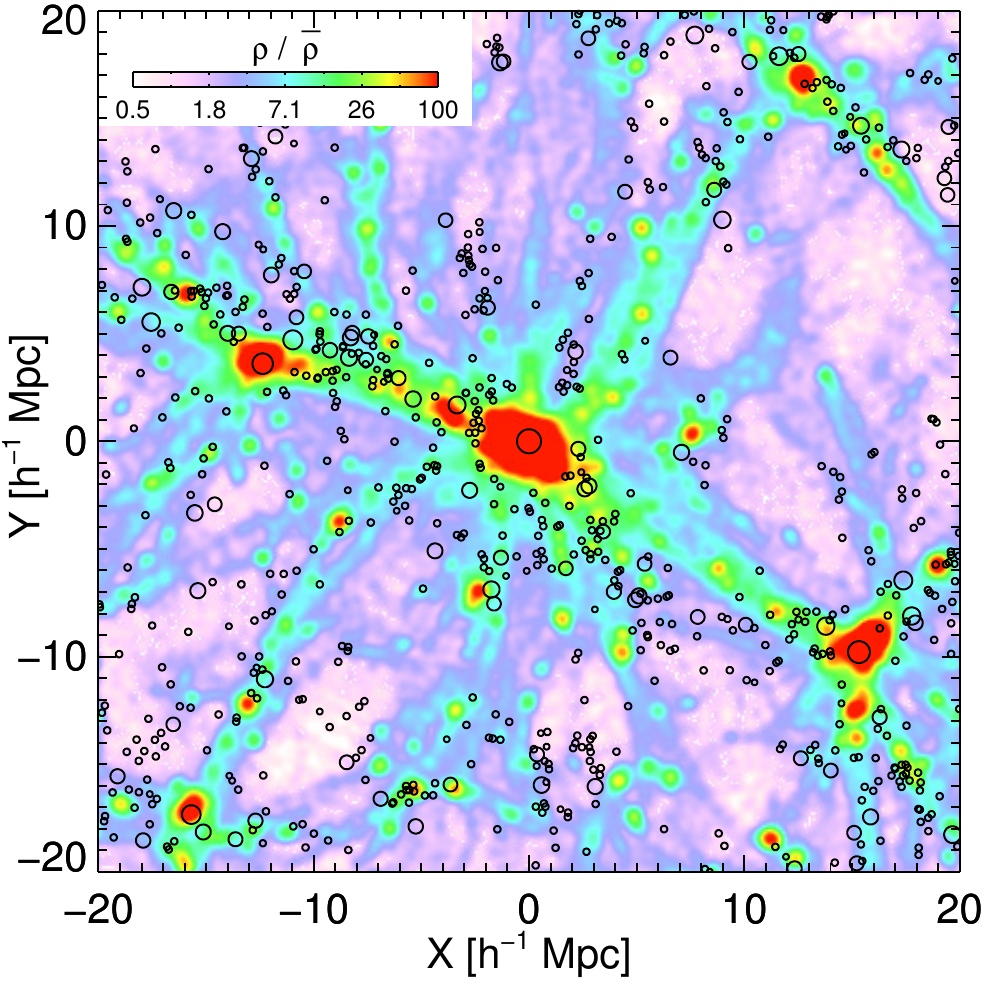}
	\caption{The projected mass density in a region of $40\mpch\times40\mpch{}$ around the simulated dark matter halo which corresponds to the Coma Cluster. The projection depth is $10h^{-1}{\rm Mpc}$. The colour-coded map shows the density field, with red and white colours indicating high and low density regions respectively (see colour bar). The black open circles indicate the observed positions of the Coma cluster and other galaxy groups around it, with sizes proportional to their estimated mass, $M_{\rm 200}$.}
	\label{fig:coma_density}
\end{figure*}

In Figure \ref{fig:coma_force} we show the fifth-force-to-standard-gravity ratio in the same region as Figure \ref{fig:coma_density}, for four different $f_{R0}$ parameter values as indicated in the legends of the four panels. As expected, in the inner regions of the cluster screening is more efficient, due to the deeper Newtonian potential there. As $|f_{R0}|$ decreases, screening becomes more efficient; for $|f_{R0}|=10^{-6}$, which is the model with the weakest screening, the fifth force is strongly suppressed (with force ratio $F_{5\rm th}/F_{\rm standard}\lesssim0.01$) only up to $\sim2~h^{-1}$Mpc from the cluster; as $|f_{R0}|$ decreases, this screened region (blue or red in colour) expands outwards, with the nearby filamentary structures and some smaller haloes scattered around now also featuring a strongly suppressed fifth force. 

\begin{figure*}
	\plotone{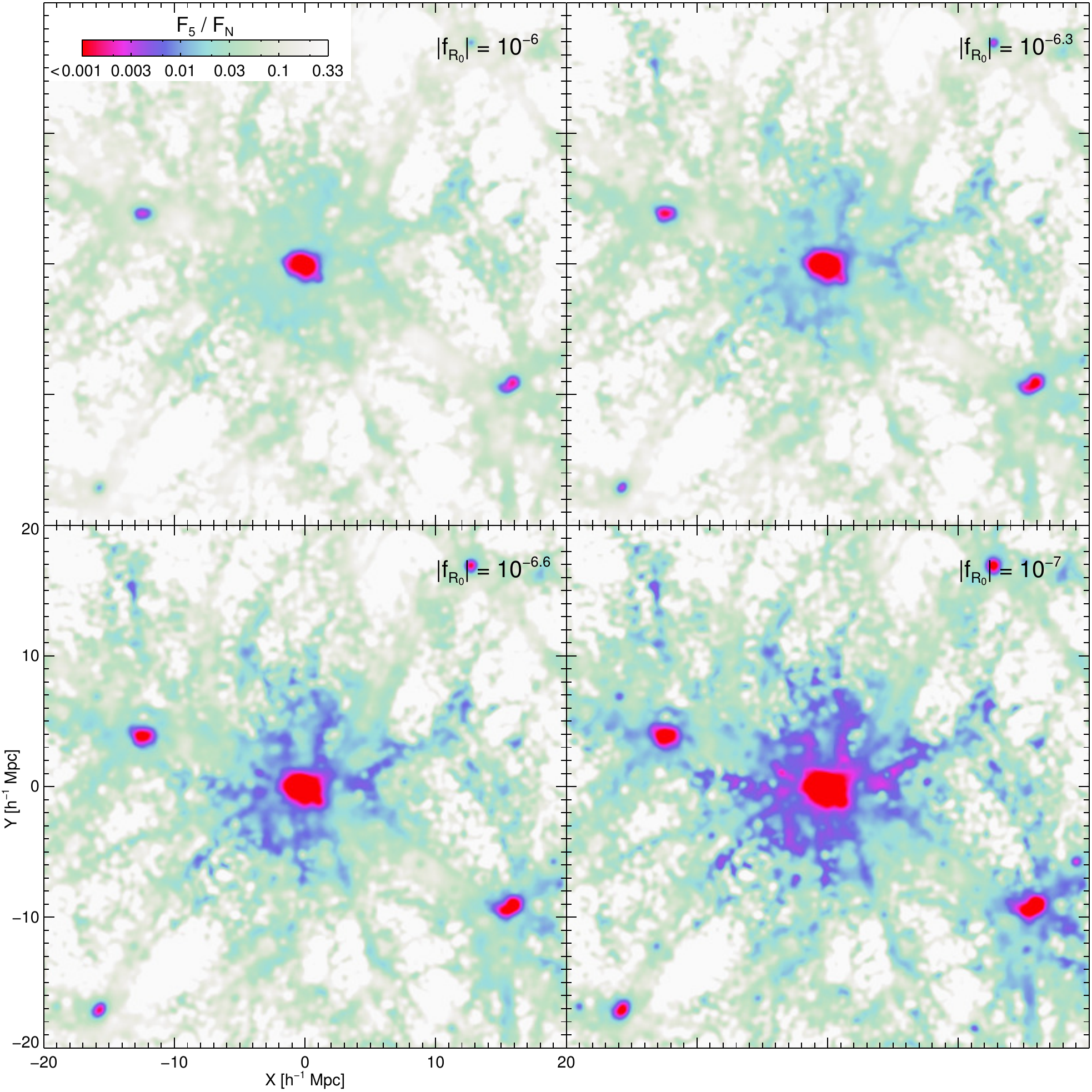}
	\caption{The fifth-force-to-standard-gravity ratio in a region of $40~h^{-1}{\rm Mpc}\times40~h^{-1}{\rm Mpc}\times10~h^{-1}{\rm Mpc}$ around the simulated dark matter halo that corresponds to the Coma Cluster. Each panels corresponds to a different value of the	$f(R)$ gravity parameter, $f_{R0}$, as indicated in the legends. The various colours indicate the median value of the force ratio of particles in each cell (see legend). Note that the cells without any particle are indicated with white colour. The figure shows that as $|f_{R0}|$ decreases, ever larger regions around the Coma cluster become screened.}
	\label{fig:coma_force}
\end{figure*}

Finally, Figure \ref{fig:coma_profile} shows the density (top panel) and force ratio (right) profiles in the Coma halo. The density profile is obtained by computing the spherically averaged densities within logarithmic radial bins from the halo centre found by {\sc rockstar}, and we show the result out to $5h^{-1}$Mpc from the halo centre, with the halo radius $R_{200}$ indicated by the dashed vertical line. The profile can be well fitted by the Navarro-Frenk-White \citep[][NFW]{Navarro:1995iw} formula,
\begin{equation}
    \rho(r) = \frac{\rho_0}{r/R_s\left(1+r/R_s\right)^2},
\end{equation}
in which $\rho_0$ is a characteristic density and $R_s$ the scale radius, and the best-fit value of $R_s$ is found to be $0.65\mpch$, so that the halo concentration is 
\begin{equation}
    c_{200} \equiv \frac{R_{200}}{R_s} = 2.3.
\end{equation}
The best-fit NFW profile for this halo is plotted as the black dotted line. {Here, we use the particles within $R_{200}$ to fit the NFW profile, and the concentration would be greater if we extend the fitting to larger radii.}

The lower panel of Fig.~\ref{fig:coma_profile} shows the force ratio profiles in the same halo for the different $|f_{R0}|$ values, decreasing from top to bottom. This is obtained similarly as the density profiles, but the spherical average is now over the force ratio at the positions of all simulation particles for each radial bin. As is typical for haloes of this mass, the fifth force is efficiently suppressed inside $R_{200}$ even for the model which deviates most from GR ($|f_{R0}|=10^{-6}$). Another interesting feature is that the shapes of the force ratio profiles are similar for all $f_{R0}$ values, and the only difference is in the amplitudes. This is a natural consequence of using the same density profile for all our fifth force calculation in all models. Note that baryonic processes associated with the cluster and galaxy evolution may also play a role in the redistribution of matter and in the suppression of the fifth force. It would be beneficial to test in a zoom-in hydrodynamics simulations, which will be left as a potential project for the future.

\begin{figure}
	\plotone{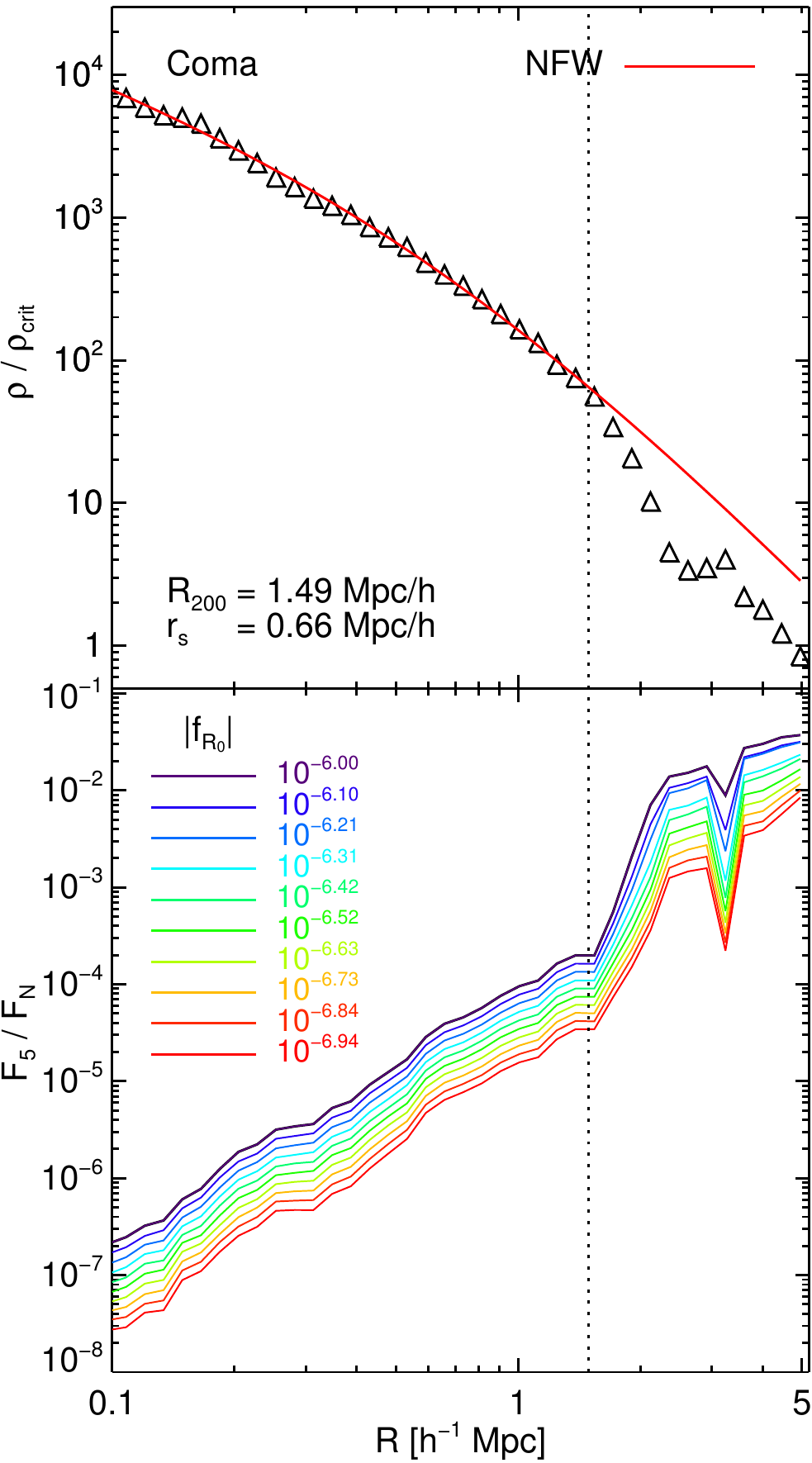}
	\caption{Top panel: the spherically averaged matter density profile as a function of the distance from centre, for the simulated dark matter halo that corresponds to the Coma Cluster. Bottom panel: the spherically averaged fifth-force-to-standard-gravity ratio profiles for the simulated halo that corresponds to the Coma Cluster, for a range of $f(R)$ gravity parameters shown by different colours, as indicated by the legend. The vertical dotted line indicates the $R_{\rm 200}$ of the halo.
	\newline
	}\label{fig:coma_profile}
\end{figure}

\section{Summary and conclusions}
\label{sect:conclusions}

We have developed a new methodology for testing modified gravity theories using astrophysical probes \citep{Jain:2013wgs,Sakstein:2018} based on constrained simulations of the local Universe. This method takes advantage of the recent developments in reconstructing the density field (and its initial conditions) of the local Universe \citep[e.g.,][]{Wang2014,Lavaux:2015tsa,Sorce:2015yna,Carlesi:2016qqp,Wang2016}, which provide a way to realistically include the environmental effects that are often important in quantitatively determining the behaviour of gravity in MG models. Our method combines the large-scale density field from these reconstruction schemes with the fully nonlinear numerical solution to the MG equations achieved through the {\sc ecosmog} code. Assuming that the matter field at the low-$z$ Universe\footnote{We have argued that this is a good approximation for most cases, but note that this approximation is not needed: full simulations with MG are possible though more time consuming.} behaves similarly in realistic MG models and $\Lambda$CDM, this will make it possible to create screening maps for a large number of MG models and parameter choices at a relatively low cost.

This is the first of a series of papers, where we have presented the methodology and, as a proof of concept, shown screening maps and some statistical properties that one can extract. As demonstrated in Figures~\ref{fig:map} and \ref{fig:galaxy}, the simulated halo distributions and the resulting screening maps show good visual agreements with the distribution of {\sc sdss} galaxies and groups, indicating that the method is capable of telling, for a given MG model, which parts of the local Universe and how well they are screened. The force behaviours displayed in Figures \ref{fig:force_ratio_z0}, \ref{fig:force_angle_z0} and \ref{fig:morphology} also agree with expectations based on the properties of chameleon screening, with smaller $|f_{R0}|$ values generally corresponding to more strongly suppressed fifth forces. In particular, Figs.~\ref{fig:force_ratio_z0} and \ref{fig:morphology} show that the fifth force is suppressed in not only high-density regions (where the Newtonian force is strong) but also in low-density regions. This seemingly counter-intuitive effect is due to the fact that the Compton wavelength of the scalar field in the models is small, such that there are few particles the fifth forces produced by which could propagate into deep voids \citep{Paillas:2018wxs}. 

As a specific example, we have analysed in greater detail the dark matter halo from our simulation box which is the counterpart to the Coma cluster. Figure \ref{fig:coma_force} shows that for all models considered here, the central region within $R\sim2~h^{-1}$Mpc is well screened and so gravity there should behave like GR. On the other hand, in the stronger screening cases, where $|f_{R0}|\rightarrow10^{-7}$, the screened region becomes larger, showing that the presence of a massive body can screen its smaller neighbours. This can be seen more clearly in the lower panel of Figure \ref{fig:coma_profile}, which shows that within the virial radius the fifth force has never exceeded $\sim0.01\%$ of the Newtonian force for all models considered.

Screening maps as shown in this paper can be invaluable for astrophysical tests \citep[e.g.,][]{Cabre:2012,Desmond2018,Desmond:2017ctk,Desmond:2018kdn,Desmond2019}, and they will enable these tests to become more reliable. However, the application of these maps in real tests are beyond the scope of this paper and will be left as future work. Also, one slight limitation of the current maps is that the Local Group is not included in the {\sc sdss} field, but this is not a practical restriction for our method considering that constrained realisations that include the Local Group have now been produced by various groups. One interesting possibility is to use such constrained initial conditions to run very-high-resolution zoom-in simulations, possibly with baryons, which realistically reproduce the basic observational properties of the Milky Way Galaxy, and use that to quantify the screening inside the Milky Way and in the Solar system.

Another point that merits further investigation is related to the uncertainty in the estimation of galaxy host halo mass. The reconstruction method used here, from \citet{Wang2016}, can effectively trace the $z=0$ massive haloes ($\gsim 10^{13.5} \Msun$) back to their initial condition, but its accuracy in recovering the matter distribution is poorer on smaller scales: the uncertainty was found to be respectively 0.23 dex on $2 \mpch{}$ and 0.1 dex on $4 \mpch{}$ scales. Therefore, in the screening map, while the environmental screening effect caused by large-scale structure can be reliably modelled, the uncertainty related to the galaxy host halo mass estimation can be another source of error, in particular for models with small Compton wavelengths of the scalar field. The impact of this uncertainty on the screening map and consequently on the error of model parameter constraints, however, can be assessed by creating a large number of screening maps using modified density fields in regions that correspond to dark matter haloes (to reflect the mass error of those haloes) before calculating the scalar field, and using these maps to quantify the scatter.

\section*{Acknowledgements}
We thank the anonymous referee for detailed comments that have helped us improve the paper. SS and BL are supported by the European Research Council (ERC) via grant ERC-StG-716532-PUNCA. MC and BL are supported by STFC Consolidated Grants ST/P000541/1, ST/L00075X/1. MC is also supported by the EU Horizon 2020 research and innovation programme under a Marie Sk{\l}odowska-Curie grant agreement 794474 (DancingGalaxies) and by the ERC grant DMIDAS [GA 786910]. HYW is supported by the National Key R\&D Program of China (grant No. 2018YFA0404503) and NSFC 11421303. JW is supported by STRD grants (2015CB857005, 2017YFB0203300) and NSFC grant 11873051. This work used the DiRAC Data Centric system at Durham University, operated by the Institute for Computational Cosmology on behalf of the STFC DiRAC HPC Facility (\url{www.dirac.ac.uk}). This equipment was funded by BIS National E-infrastructure capital grant ST/K00042X/1, STFC capital grants ST/H008519/1, ST/K00087X/1, STFC DiRAC Operations grant ST/K003267/1 and Durham University. DiRAC is part of the National E-Infrastructure.

\vspace{-.0cm}
\bibliographystyle{mnras}
\bibliography{bibliography}
\label{lastpage}
\end{document}